\documentclass[prb,notitlepage,twocolumn,superscriptaddress]{revtex4}
\usepackage{amsmath,amssymb,color,graphicx,hyperref,amsthm,bm,verbatim}

\newcommand{\rescaler}{\mathfrak{r}}

\begin{document}

\title{Hierarchies of Length Scale Based Typology in Anisotropic Multiband Superconductor}

\author{Thomas Winyard}
\affiliation{School of Mathematics, University of Leeds, Leeds LS2 9JT, United Kingdom}
\affiliation{Department of Physics, KTH-Royal Institute of Technology, Stockholm, SE-10691 Sweden}
\author{Mihail  Silaev}
 \affiliation{Department of
Physics and Nanoscience Center, University of Jyv\"askyl\"a, P.O.
Box 35 (YFL), FI-40014 University of Jyv\"askyl\"a, Finland}

\author{Egor Babaev}
\affiliation{Department of Physics, KTH-Royal Institute of Technology, Stockholm, SE-10691 Sweden}

\begin{abstract} 
  Since Ginzburg and Landau's seminal work in 1950 superconducting states have been classified by the hierarchy of the fundamental length scales of the theory; the magnetic field penetration lengths and coherence lengths. In the simplest single-component case they form a dimensionless ratio $\kappa$. 
{The model  was generalized by Ginzburg for anisotropic materials in 1952. 
  %
In this paper we expand the above length scale analysis to anisotropic multi-component superconductors, that can have multiple coherence lengths as well as multiple magnetic field penetration lengths, leading to unconventional length scale hierarchies. 
We demonstrate that the anisotropies in multi-band superconductors lead to new regimes with various mixed hierarchies in different directions. For example, a regime is possible, where for a field applied in a certain direction coherence lengths are smaller than the magnetic field penetration lengths in one of the perpendicular directions, where as the penetration lengths are larger in the other direction.  Examples are shown of a new regime where vortex cores overlap in one direction, resulting in attractive core-core interaction, while in the orthogonal direction the magnetic field penetration length exceeds the coherence lengths leading to dominance of repulsive current-current interaction resulting in an
unconventional magnetic responce.}

\end{abstract}

\maketitle
\section{Introduction}
The goal of this paper is to calculate and classify the coherence and magnetic field penetration length hierarchies and their effects on the magnetic properties of anisotropic multiband superconductors.
The original Ginzburg-Landau theory \cite{landau1950k} classified superconductors by a single number, the Ginzburg-Landau parameter $\kappa=\lambda/\xi$, constructed from the ratio of the two fundamental length scales of the classical Ginzburg-Landau effective field theory; magnetic field penetration length $\lambda$ and coherence length $\xi$.
Using this framework they identified two regimes in an externally applied magnetic field; $\kappa=\lambda/\xi<1$ where the free energy of a transnationally-invariant superconductor-to-normal interface was positive and $\kappa=\lambda/\xi>1$ where the interface energy was negative [in the original paper the boundary (or critical) value of $\kappa$ was $1/\sqrt{2}$, we have absorbed the factor $\sqrt{2}$ into the definition of the coherence length]. Later it was firmly established that for $\kappa=1$, any arbitrarily shaped superconductor-to-normal metal interface in critical magnetic field has zero energy \cite{saint1969type,kramer,Bogomol,bogomol1976stability}.
{Almost immediately after the formulation of the theory, two important generalisations were discussed.
Superconducting materials are in general anisotropic, with coherence and penetration lengths having a directional dependence. 
A corresponding generalization of the Ginzburg-Landau effective theory was discussed in \cite{ginzburg1952,gor1964microscopic,tilley1965ginsburg,tilley1965critical,kats1969some}.
Shortly after the formulation of Bardeen-Cooper-Schrieffer theory, the superconducting state was generalized to various non-$s$ paring states. It was demonstrated that superconducting states in general break multiple symmetries and are
thus described by multicomponent Ginzburg-Landau theory.
Multiple components could originate from several superconducting components in different bands, even without the breaking of multiple symmetries by the superconducting state. Namely, in \cite{Suhl1959,Moskalenko59superconductivityof} superconductivity models were generlized to the case of multiple bands. A microscopic derivation of the two-band generalization of the Ginzburg-Landau model (using two complex fields), corresponding to superconductivity in different bands, followed shortly \cite{Tilley1964} \footnote{At this time, the multiband generalization of the Ginzburg-Landau expansion was not rigorously justified. This was due to the systems that were considered \cite{Tilley1964} not admitting an expansion in several order parameters, due to the $U(1)$ symmetry, as in the single-band model. Therefore, such an expansion cannot be mistaken for an expansion in a single (small) parameter $\tau=(1-T/T_c)$, similar to any systems with explicitly broken symmetries. In the fermionic pairing context, this problem of justifiability of a Ginzburg-Landau field theory, in systems with explicitly broken symmetries, was later understood in  other context  \cite{nambu1961dynamical} and formally justified in connection with a $U(1)$ multiband system \cite{Silaev2012a}. Recently mathematical derivations of multicomponent expansions, yielding multiple coherence lengths, have been obtained \cite{frank2016multi}}. 
}

{ The new regime that is possible in multicomponent isotropic systems, compared to their  single-component counterparts, originates from some of the coherence lengths being smaller and others larger than the magnetic field 
} penetration length \cite{Babaev.Speight:05, Babaev.Carlstrom.ea:10,Silaev.Babaev:11,Carlstrom.Garaud.ea:11a,nonpairwise}. I.e. in the $n$-component case   $\xi_1, < \xi_2< ...\lambda...<\xi_{n-1}<\xi_n$. This regime has been termed ``type-1.5" in the experimental paper \cite{Moshchalkov.Menghini.ea:09}. 
 For artificially layered systems, the case of more complicated length scale hierarchies, stemming from different penetration lengths in different layers, were considered \cite{varney2013hierarchical,diaz2017glass}.

{Hierarchy of magnetic length scales was classified at the level of London model for anistropic systems.
A new feature that arises in single-component anisotropic model is the appearance of two magnetic modes associated with the different polarizations of the magnetic field, leading, under certain conditions to field inversion \cite{Kogan1981,Buzdin,Balatskii}. In multi-band anistropic models the situation is more involved.
In  \cite{silaev2018non} it was shown that qualitatively different  electromagnetic effects arise when both anisotropy and multiple bands are present. Namely it was demonstrated that a London model with $n$ bands with different  anisotropies, the magnetic mode 
hybridizes with Leggett's modes and as the result the systems have in general $n+1$ magnetic modes with different magnetic field penetration lengths. The magnetic field penetration is characterized by $n+1$ exponents that are different in different directions and under certain conditions have oscillatory behaviour. 

This calls for investigation of these superconducting regimes including interplay between the coherence 
and magnetic field penetration lengths which requires going beyond the London models. 
 In the present paper we present such an analysis by considering the Ginzburg-Landau model of the 
multiband anisotropic superconductors. That is, we discuss the most general situation when both the magnetic field is characterised by multiple penetration depths and there are several distinct and directional-dependent coherence lengths. 
}

The outline of this paper is the following. 
First we analyse the normal modes and calculate the coherence lengths for a given anisotropic multiband model. 
We then analyse and classify possible hierarchies of the length scales. 
Then we discuss the implications of this for vortex solutions, the magnetic response of the system and discuss asymptotic intervortex forces for simple inter-band couplings. After that we focus on multivortex solutions and the magnetic response in regimes that are not present in the isotropic counterpart of the model, nor in the London limit of the multiband model considered in \cite{silaev2018non}.
    
 \section{The Model}
 We consider the multiband Ginzburg-Landau free energy for an 
 $n$-component anisotropic system given by the free energy 
 { in dimensionless units}, 
 \begin{equation}
 F =\frac{1}{2}\int_{\mathbb{R}^3}\left\{ \sum^{n}_{\alpha = 1} \left(\gamma^{-1}_{ij\alpha}D_j\psi_{\alpha}\right)
 \left(\gamma^{-1}_{ik\alpha}  \overline{D_k\psi_{\alpha}}\right) + \boldsymbol{B}^2 + F_p \right\},
 \label{GL}
 \end{equation}
 \noindent where $D_i = \partial_i + ieA_i$ is the covariant derivative and $\psi_\alpha = \left|\psi_\alpha\right|e^{i\theta_\alpha}$ represents the different superconducting components, that for example can be superconducting components in different bands. Greek indices will always be used to denote superconducting components and Latin indices will be spatial, with the summation principle applied for repeated Latin indices only. The anisotropy of the system is given by $\gamma_{ij\alpha}$ which represents a 3 dimensional diagonal matrix for each component,
 \begin{equation}
 \gamma_{ij\alpha} = \left(\begin{array}{ccc} \gamma_{x \alpha} & & 
 \\ & \gamma_{y \alpha} & 
 \\ & & \gamma_{z \alpha} \end{array} \right).
 \label{Eq:matrices}
 \end{equation}
 \noindent $F_p$ collects together the potential (non-gradient) terms which can be any from a large range of gauge invariant terms. The simplest example, and the one we will mostly focus on, is the standard situation of a clean $s$-wave multiband superconductor, with the potential terms and the Josephson-Leggett inter-band coupling term.
 \begin{eqnarray}
 F_p &=& \sum^n_{\alpha = 1}\frac{\Gamma_{\alpha}}{4}\left({\psi^0_\alpha}^2 - \left|\psi_\alpha\right|^2\right)^2  
 \nonumber 
 \\ && 
 - \sum^n_{\alpha  = 1}\sum_{\beta < \alpha}\eta_{\alpha \beta} \left|\psi_\alpha\right|\left|\psi_\beta\right| 
 \cos{\left(\theta_{\alpha\beta}\right)},
 \label{Eq:pot}
 \end{eqnarray}
where ${\psi^0_\alpha}$, $\Gamma_{\alpha}$ and $\eta_{12}$ are positive real constants. The second term above is the Josephson inter-band coupling, where $\theta_{\alpha\beta} = \theta_\alpha - \theta_\beta$ is the inter-band phase difference between components $\alpha$ and $\beta$. We especially focus here on the case where the Josephson term locks all phase differences to zero in the ground state, thus explicitly breaking the symmetry from $U(1)^n$ to $U(1)$.  
For a detailed discussion of the microscopic justification of this kind of multiband Ginzburg-Landau expansion see \cite{Silaev.Babaev:12}. 

{ Furthermore we focus on the two-dimensional case when the magnetic field depends on the two coordinates
 $\bm B = \bm B(x,y)$. Besides that we consider the tetragonal crystal symmetry and assume that the 
 field is directed along one of the crystal axes, to be definite $\bm B = B\bm z$.  
 In this case there appear no additional components of the field. 
 Note that for such magnetic field direction  the anisotropy can  
 be removed by a suitable spatial rescaling, which has the effect of merely rescaling some of the parameters in the model.
 However in the two-band model such rescaling is not possible, provided the matrices in \ref{Eq:matrices} are linearly independent of each other for each order parameter field component.} By rescaling coordinates it is now only possible to isotropise one of the bands, while the others remain anisotropic e.g. for the 1st band $x_i \rightarrow \gamma_{1ij}^{-1}x_j$ and $ A_i \rightarrow \gamma_{1ij} A_j$.

 In studying the magnetic response, we primarily focus on composite vortices (winding in each component is equivalent) since fractional vortices have infinite energy in a bulk sample \cite{frac,Silaev:11}. Hence we can categorise each solution by the winding number $N$ of the complex phase of both of the condensates. $N$ also dictates the magnetic flux through the plane which is quantised,
\begin{equation}
N = \frac{\Phi}{\Phi_0} = \frac{1}{2\pi}\int_{\mathbb{R}^2} B d^2 x,
\end{equation}
where $\Phi_0$ is the flux quantum. As discussed the anisotropy in general breaks the spatial symmetries to lower subgroups, however there are a few special choices of parameters, namely the isotropic case which would lead to the familiar $O(2)$ spatial symmetry in the 2-dimensional model and also choices that lead to higher $D_4$ dihedral symmetries for even bands (combined with an interchange of the fields $\psi_\alpha \rightarrow \psi_\beta$).

We have previously demonstrated that physics, which  has no counterpart in the single-component models, arises  when the anisotropy in each band is not equivalent. This condition leads to multiple magnetic field penetration lengths in the London limit and, under certain circumstances, to a non-local electromagnetic response in the nominally local London model \cite{silaev2018non}. {Importantly, the non-locality scale in this case is determined  by the strength of inter-band Josephson coupling and has nothing to do with the non-localities of the usual BCS theory, associated with the Cooper pair dimension \cite{Bardeen1957,Bardeen1957a}. }

These unusual electromagnetic properties lead to the possibility of different length scale hierarchies in different directions for anisotropic multi-band superconductors. As established in the London model \cite{silaev2018non} the system has multiple magnetic field penetration lengths.
Therefore the new possible hierarchies are:
\begin{itemize}
\item
In one direction $(\hat{{\bf r}})$ the system has all coherence lengths smaller than all penetration lengths $\xi_1(\hat{{\bf r}}),\xi_2(\hat{{\bf r}}),...,\xi_n(\hat{{\bf r}})>\lambda_1(\hat{{\bf r}}),\lambda_2(\hat{{\bf r}}),...\lambda_{n+1}(\hat{{\bf r}})$ (type-1)
\item
in another direction all penetration lengths are smaller than the coherence lengths $\xi_1(\hat{{\bf r}}),\xi_2(\hat{{\bf r}}),...,\xi_n(\hat{{\bf r}})<\lambda_1(\hat{{\bf r}}),\lambda_2(\hat{{\bf r}}),...\lambda_{n+1}(\hat{{\bf r}})$ (type-2)
\item
in some directions the hierarchy is mixed, i.e. some penetration length(s) $\lambda_i(\hat{{\bf r}})$
are smaller and some are larger than the coherence lengths $\xi_i(\hat{{\bf r}})$ (type-1.5)
\end{itemize}

In an isotropic multiband superconductor, one of the consequences (although not a state-defining one) of different length scale hierarchies is the following: interactions between two vortices with similar phase windings is isotropically attractive in the case
where all coherence lengths are larger than the penetration length, due to 
domination of core-core interaction. 
Conversely it is repulsive in the case where the magnetic field penetration length is larger than coherence lengths,
unless there is a field inversion.
In the case where magnetic field penetration length falls between the coherence
lengths the inter-vortex interactions are attractive at longer and repulsive at shorter range (see \cite{Babaev.Speight:05,Silaev.Babaev:11,Carlstrom.Garaud.ea:11a,nonpairwise,babaev2017type} ).
In the multiband anisotropic case, the fact that anisotropies cannot be rescaled { even for the fields
directed along the crystal axes}, suggests that the typology of superconductivity states require specifying 
{  length scale hierarchies for different directions} in a plane. That is, since the hierarchy of the fundamental length scales are different in different directions, intervortex interactions in one direction can be dominated by core-core intervortex forces and in another direction by electromagnetic and current-current interaction. 
To find the range of parameters where such effect can take place we
consider the linearised theory. Then, to find the actual  vortex configurations 
we need to use numerics because of the highly complex and non-linear nature of the problem of investigating these vortex states. 
  
\section{Single-Vortex Solutions}
 To understand the basic properties of the vortex states in anisotropic multiband superconductors we consider first the single quanta $N=1$ solutions. 

For numerical calculations we use the FreeFem++ library on a finite element space. A conjugate gradient flow method was utilised to minimise various initial conditions to find the minimum that is displayed. All initial  configurations  took the form of perturbed spherically symmetric vortices, either with higher winding number or well separated such that they can still interact in a reasonable time.  The grid dimensions where chosen to be substantially larger than the scale of the vortices, such that vortices do not interact with boundaries.  We have considered many parameters in our investigation and have selected some particular choices that demonstrate the key behaviour of the systems we are interested in.

The initial conditions used for introducing both single and multi quanta vortices, when well separated, can be written $\psi_\alpha = \left(\prod_{k=1}^{N} \psi_{\alpha}^{(k)}(\boldsymbol{x})\right)/u_\alpha^{N-1}$, where $u_\alpha$ is the ground state (vacua) value for the magnitude and the radial ansatz for a vortex at the origin is $\psi^{(k)}_\alpha(0) = f_\alpha(r) e^{i \theta}$. The profile functions have the limits $f_\alpha(0) = 0$, $f_\alpha(\infty) = u_\alpha$.

\begin{figure}[tb!]
 \includegraphics[width=1.0\linewidth]{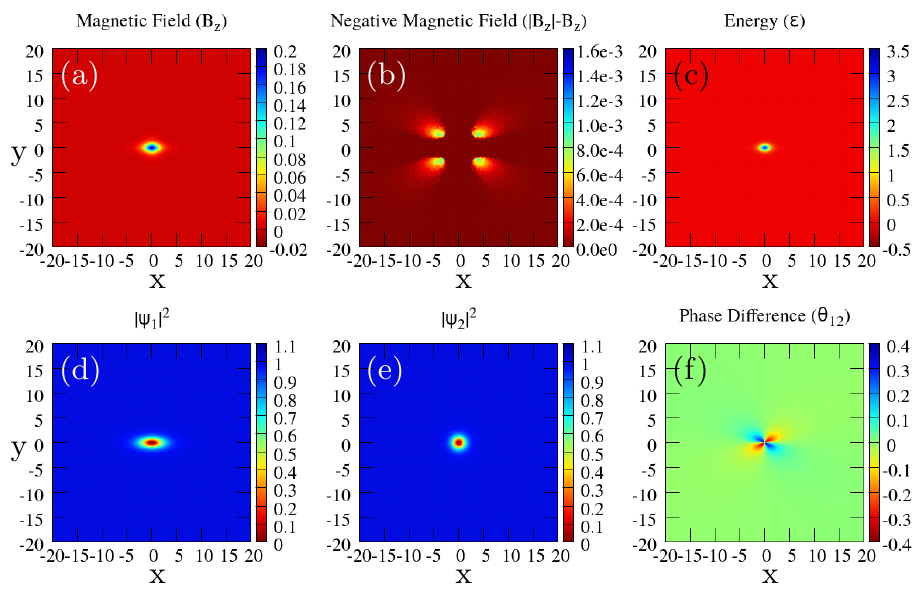}    
 \caption{\label{Fig:charge1} (Colour online) 
 $N=1$ one quanta numerical solution for anisotropy in one band with no Josephson coupling $\Gamma_1=\Gamma_2=2$, $\eta_{12}=0$, $\gamma^{-1}_{1x}=2, \gamma^{-1}_{1y}=\sqrt{0.7}$ and $\gamma_{2x}=\gamma_{2y} = 1$. (a) $B_z$ magnetic field, (b) $\left|B_z\right| - B_z$ negative magnetic field, (c) $\mathcal{E}$ energy density, 
 (d) $\left|\psi_1\right|^2$, (e) $\left|\psi_2\right|^2$, (f) $\theta_{12}$ phase difference.}
 \end{figure}

\begin{figure}[tb!]
 \includegraphics[width=1.0\linewidth]{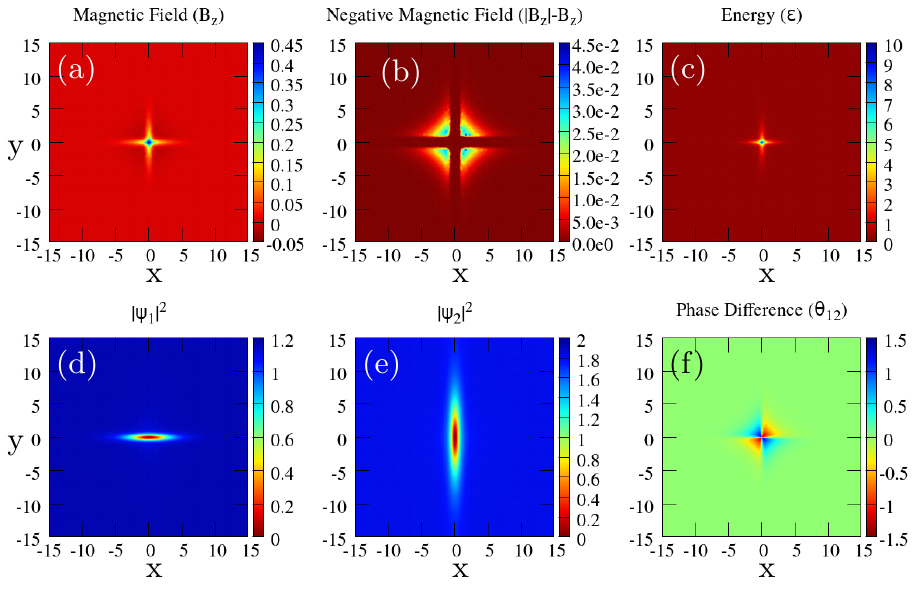}    
 \caption{\label{Fig:charge1strong} (Colour online) 
 $N=1$ one quanta numerical solution for very strong anisotropy in opposite bands with Josephson coupling and different parameters $\Gamma_1=4, \Gamma_2=0.5$, $\eta_{12}=1$, $\gamma^{-1}_{1x}=\gamma^{-1}_{2y} = 4, \gamma^{-1}_{1y}=\gamma^{-1}_{2x}=0.5$. (a) $B_z$ magnetic field, (b) $\left|B_z\right| - B_z$ negative magnetic field, (c) $\mathcal{E}$ energy density, 
 (d) $\left|\psi_1\right|^2$ (e) $\left|\psi_2\right|^2$,  (f) $\theta_{12}$ phase difference.}
 \end{figure}

\begin{figure}[tb!]
 \includegraphics[width=1.0\linewidth]{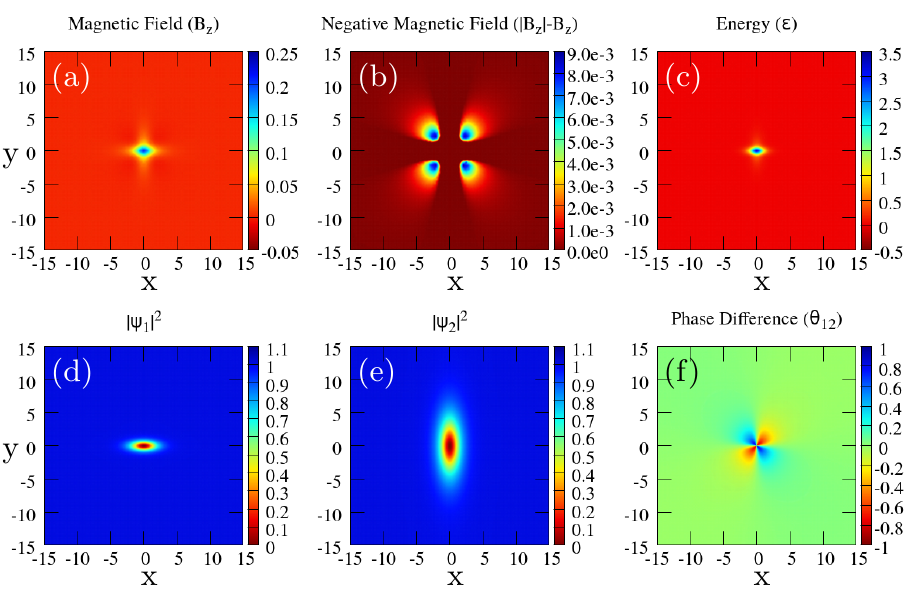}    
 \caption{\label{Fig:charge1strongnok} (Colour online) 
 $N=1$ one quanta numerical solution for anisotropy in both bands with no Josephson coupling $\Gamma_1=4, \Gamma_2=0.5$, $\eta_{12}=0$, $\gamma^{-1}_{1x}=\gamma^{-1}_{2y} = 2, \gamma^{-1}_{1y}=\gamma^{-1}_{2x}=\sqrt{0.7}$.
  (a) $B_z$ magnetic field, (b) $\left|B_z\right| - B_z$ negative magnetic field, (c) $\mathcal{E}$ energy density, 
 (d) $\left|\psi_1\right|^2$, (e) $\left|\psi_2\right|^2$, (f) $\theta_{12}$ phase difference.}
 \end{figure}

Using the above  {  initial guess}, along with perturbations to ensure the radial symmetry is broken, we minimise using the conjugate gradient flow algorithm to find the true minimal energy solutions for single quanta, similar to those displayed in figures \ref{Fig:charge1}, \ref{Fig:charge1strong} and \ref{Fig:charge1strongnok}. From these solutions it is clear that the hierarchy of the length scales associated with matter fields and magnetic fields can be different in different directions.
 
As with the strong type-2 case \cite{silaev2018non}, we observe field inversion (negative magnetic field). Additionally we observe that the field inversion is present when the Josephson coupling is set to zero ($\eta_{12} = 0$) in figure \ref{Fig:charge1strongnok}, but also that the mode  that mediates the negative magnetic field become more long range but also weaker.

The main conclusion that can be drawn from the single-vortex solutions presented in this section is that the core can extend beyond the flux-carrying area in certain directions while being smaller than the magnetic field localization in other directions, as seen in figure \ref{Fig:charge1strong}. 
 
 
\section{ Vortex interaction energies within the linearised theory }
 In this section we consider the fundamental length scales by performing  
 an asymptotic analysis and linearising the Ginzburg-Landau model near its 
 ground state. We consider a system where the global minima occurs at 
 $\left(\left|\psi_\alpha\right|,\theta_{\alpha\beta}\right) = \left(u_
 \alpha,0\right)$ and $u_\alpha \geq 0$. The trivial ground state solution  
 is then given to be $\psi_\alpha = u_\alpha$, $\boldsymbol{A} = 0$ 
 and $\theta_{\alpha\beta} = 0$.  

 As we are interested in asymptotic interactions between vortices, we 
 { consider leading-order terms in the free energy which are quadratic in the small fluctuations} of the vector field  $\boldsymbol{A}$ and the real scalar fields $\epsilon_\alpha = \left|\psi_\alpha\right|-u_\alpha$ and 
 $\theta_{12}$. 
 While for asymptotic inter-vortex forces in the standard isotropic $s$-wave multiband superconductor it can be assumed that $\theta_{12} = 0$ \cite{Carlstrom.Babaev.ea:11,Babaev.Carlstrom.ea:10}, this is no longer the case for anisotropic systems, as gradients of the phase difference and the magnetic field are coupled due to the anisotropy \cite{silaev2018non}. 
 The general leading-order terms in free energy can be calculated as follows,
 \begin{widetext}
 \begin{eqnarray}
 F_{lin} &=&\sum_{\alpha = 1}^{n}\left(\gamma_{\alpha}^{-2}\nabla\epsilon_ 
 \alpha\cdot\nabla\epsilon_\alpha\right) + \frac{1}{2}\sum^n_{\alpha = 1}
 \sum_{\beta = 1}^{n} \epsilon_\alpha \mathcal{H}_{\alpha\beta} \epsilon_
 \beta + \sum^n_{\alpha = 1} (\hat \gamma_\alpha^2 {\bm j_\alpha}\cdot{\bm 
 j_\alpha} )+ \frac{1}{2}\left(\partial_1 A_2 - \partial_2 A_1\right)^2 
 + \sum^n_{\alpha = 1}\sum_{\beta < \alpha} J_{\alpha\beta} \theta_{\alpha
 \beta}^2.
 \label{Eq:linear}
 \end{eqnarray}
 \end{widetext}
 \noindent Here $\bm{j}_\alpha$ are the partial superconducting currents
 \begin{equation}\label{Eq:Currents}
 {\bm j}_{\alpha}=\frac{1}{2e}\hat\gamma_{\alpha}^{-2}
 \left( \boldsymbol{\nabla}\theta_{\alpha}- e{\bm A} \right),
 \end{equation}
where have set $\hbar = c = 1$, $\hat\gamma_{k}$ are coefficients characterizing the contribution of each band to the Meissner screening, ${\bm A}$ is the vector potential and $J_{\alpha\beta}$ the Josephson coupling. Additionally $\mathcal{H}$ is the Hessian of the potential term $F_p \left(\left|\psi_\alpha\right|\right)$ evaluated about the field vacuum values $\left(u_\alpha\right)$,
\begin{equation}
\mathcal{H}_{\alpha\beta} = \left.\frac{\partial^2 F_p}{\partial \left|\psi_\alpha\right| \partial \left|\psi_\beta\right|}\right|_{\left(u_\alpha\right)}.
\label{Eq:Hessian}
\end{equation}

If we consider the linear free energy in \ref{Eq:linear}, the vector potential $\boldsymbol{A}$ and the phase 
difference $\theta_{\alpha\beta}$ are decoupled from the magnitudes of the scalar fields $\left|\psi_\alpha\right|^2$. 
Due to this decoupling we can see that \ref{Eq:linear} is split into two independent parts. That is the London-like energy is 
given by the last three terms and the first two terms yield the contribution of scalar fields. 
The London-like part of the free energy has rescaled parameters which are related to that of the Ginzburg-Landau functional Eq.(\ref{GL}) by  $J_{\alpha\beta} = \eta_{\alpha\beta} u_\alpha u_\beta$ and $ \hat\gamma_{\alpha}^{-2} = (2 e u_\alpha)^2 \gamma_\alpha^{-2} $, where the spatial matrix indices are suppressed for $\hat \gamma$ in \ref{Eq:Currents}. Note that due to being rescaled by the $u_\alpha$, if the vacuum value for the condensate magnitudes is not known analytically (for example in the case of non-zero Josephson coupling) then $\hat\gamma_{\alpha}$ also can't be known analytically.
 
 {As noted above, the magnetic and phase difference modes decouple from the condensate magnitudes and form an alternative London model, that has already been considered in \cite{silaev2018non}. For completeness we reproduce here the solutions to the linearised alternate London equations from that paper. Ultimately if we reduce to the simplest case of $n=2$, the London model leads to multiple modes being produced for both the magnetic field $\bm B$ and the phase difference $\theta_{12}$. }
In the London model  \cite{silaev2018non}, these modes take the form,  %
 \begin{align}\label{Eq:BZFin}
 & B_z (r,\varphi) =  \Phi_0\left( h_1(\varphi) \frac{e^{-k_1 r}}{\sqrt{k_1r}} - h_2(\varphi) \frac{e^{-k_2 r}}{\sqrt{k_2r}}
 \right), \\ \label{Eq:hj}
 & h_j (\varphi) =  \frac{ k_j^2 (\hat\gamma_{1x}^{-2}\hat\gamma_{2x}^{-2}\hat\gamma_{Lx}^{2}\cos^2\theta +
 \hat\gamma_{1y}^{-2}\hat\gamma_{2y}^{-2}\hat\gamma_{Ly}^{2}\sin^2\theta )- k_0^2}{a(k_1^2 - k_2^2)},
 \end{align} 
\noindent where the $k$'s are merely the wavevectors that 
are produced from the Fourier transform used to find the solution and $r$,$\theta$ represent the physical space in polar coordinates and $\hat\gamma_{Li}^2 = \left(\sum_\alpha \hat\gamma_{\alpha i}^{-2}\right)^{-1}$. 
 \begin{equation}\label{Eq:kSolution}
k_{1,2}^2 = \frac{-b \pm \sqrt{b^2 - 4ac}}{2a} ,
\end{equation}
where,
 \begin{align}
 & a = \left(\hat\gamma_{1y}^{-2} \cos^2 \theta + \hat\gamma_{1x}^{-2} \sin^2 \theta\right)\left(\hat\gamma_{2y}^{-2} \cos^2 \theta + 
 \hat\gamma_{2x} ^{-2} \sin^2 \theta\right) ,
  \\
 & b = \hat\gamma_{1y}^{-2}\hat\gamma_{2y}^{-2}\hat\gamma_{Lx}^{-2} \cos^2 \theta + 
 \hat\gamma_{1x}^{-2}\hat\gamma_{2x}^{-2}\hat\gamma_{Ly}^{-2} \sin^2  \theta 
 \nonumber \\
 & + k_0^2\left(\hat\gamma_{Ly}^{-2} \cos^2 \theta + \hat\gamma_{Lx}^{-2} \sin^2 \theta\right),\\
 & c = k_0^2\hat\gamma_{Ly}^{-2}\hat\gamma_{Lx}^{-2}.
 \end{align}
\noindent Note that there is a four fold symmetry in equation \ref{Eq:BZFin} around the vortex solution, which matches the maximal general symmetry of the original free energy.

Comparing with the numerical results of the previous section, one can see that  distributions of magnetic field and  the  phase difference $\theta_{12}$ 
are qualitatively similar to that of the London model prediction. 
Thus we suggest to use the pattern of London model solution to approximate the field obtained  in the full non-linear model,  when matched with a general profile function $b(r)$ which is a monotonic function running from $b(0) = 1$ to $b(\infty) = 0$:
\begin{equation}
B = b(r)\Phi_0\left( h_1(\varphi) \frac{e^{-k_1 r}}{\sqrt{k_1}} - h_2(\varphi) \frac{e^{-k_2 r}}{\sqrt{k_2}}
 \right).
\end{equation}

However the London model symmetries say little about the behaviour of the scalar fields (condensate magnitudes). 
 In the linearised theory  the order parameter amplitudes decouple from the magnetic field and phase difference in this model. 
From the numerical solutions obtained for the single-quantum vortex in the previous section we get that the amplitudes can be well descried by the 
axially-symmetric solutions in the rescaled coordinates  $x_i \rightarrow \gamma_{ij\alpha} x_{j}$,
\begin{equation} \label{Eq:ansatzVort}
\psi_1 = f_1\left(\rescaler_1\right)e^{i\theta_1}, \quad \quad \psi_2 = f_2\left(\rescaler_2\right)e^{i\theta_2},
\end{equation}
where $\rescaler_\alpha = \sqrt{\gamma_{\alpha x}^2 x^2 + \gamma_{\alpha y}^2 y^2}$.  Here  $f_\alpha (\rescaler_\alpha)$ are real profile functions with the boundary values $f_\alpha(0) = 0, f_\alpha(\infty) = u_\alpha$ which is dependent on the potential $F_p$. 
We expect that this approximation is accurate when inter-condensate couplings do not dominate. However when used in numerics even in the regime with strong Josephson coupling it provides very good initial guess for the solutions. 

The phase of each of the condensates in Eq.(\ref{Eq:ansatzVort}) can be written $\theta_1 = \left(\theta_{\Sigma} + \theta_{12}\right)/2 $ and $\theta_2 = \left(\theta_{\Sigma} - \theta_{12}\right)/2 $, where $\theta_\Sigma = \theta_1 + \theta_2$. 
The phase difference symmetry is determined in a similar way to the magnetic field above from the London model, equipped with an additional profile function. The phase sum however is determined entirely by the gauge choice we make along with the chosen winding number. 
\footnote{Note that while the symmetries are similar for larger Josephson coupling, but when Josephson coupling is absent, the London model predicts no additional magnetic field mode and hence no magnetic field inversion. However in the full model these features remain, due to non-linear effects ( cf \cite{Babaev.Jaykka.ea:09}). }

Here we focus on the  effects which appear beyond the London model due to the 
spatial variations of density fields, i.e. at the scales determined by  coherence lengths and  especially on their effect on the vortex states. 
To that end, for the model in question, one should analyse $\mathcal{H}$ the Hessian of the potential term $F_p \left(\left|\psi_\alpha\right|\right)$, which appears in eq. \ref{Eq:linear} (for a detailed discussion in the istropic counterpart of the model see \cite{Carlstrom.Babaev.ea:11,babaev2017type}) . 
 The remaining part of the free energy, once the decoupled London part is dealt with, is dependent on the magnitude of the scalar fields alone. In general these scalar fields $\epsilon_\alpha$ are coupled through the Hessian $\mathcal{H}_{\alpha\beta}$. This coupling can be simplified by diagonalising the Hessian in \ref{Eq:Hessian} 
 {using it's eigenvectors $v_\alpha = \left(v_\alpha^1, v_\alpha^2\right)^T$ and corresponding eigenvalues $\mu_\alpha$.} This leads to a linear combination of the fields $\chi$ where $\left(\epsilon_1, \epsilon_2\right)^T = \chi_1 v_1 + \chi_2 v_2$. In the isotropic case this would lead to a simple linear PDE for each decoupled field $\chi$. However the non-trivial anisotropy leads to additional cross terms which have the form of gradient couplings between the  fields $\chi_\alpha$. Note that these fields cannot in general be thought of as the individual condensate amplitudes. 
 Then the part of free energy which depends on the order parameter amplitude fields $\chi_{1,2}$  becomes,
 \begin{equation}
 F_{lin}^{con} = \frac{1}{2}\sum_{\alpha=1}^2 \left(\tilde{\gamma}_{\alpha i j}^{-1} \partial_j\chi_\alpha \tilde{\gamma}_{\alpha i k}^{-1} \partial_k\chi_\alpha + Q_{ij}\partial_{i} \chi_1 \partial_{j} \chi_2 + \mu_\alpha^2 \chi_\alpha^2 \right)
 \label{conlin}
 \end{equation}
\noindent where $Q_{ij} = (v_1^1 v_2^1 \gamma^{-2}_{1ij} + v_1^2v_2^2 \gamma^{-2}_{2ij})$ which vanishes in the isotropic case, due to $v_1\cdot v_2 = v_1^1 v_2^1 + v_1^2 v_2^2 = 0$ and $\tilde{\gamma}_{1}^{-2} = \gamma_2^{-2}$.

The presence of tensor $Q_{ij}$ in the anisotropic case leads to rather involved expressions for length-scales  in general, even in the simplest minimal model of a  two-band superconductor Eq. \ref{Eq:pot}. 
%
 In order to  consider the technically simplest case it is instructive to focus on the $U(1)^n$ model, namely the regime of zero Josephson coupling $\eta_{12}=0$. 
 In this case the interband phase difference degree of freedom becomes massless and decouples from the magnetic field. In this case we have only a single magnetic field although directional-dependent penetration length $\lambda_L$.
Later, we  will return the case of finite Josephson coupling.
If we return to a general number of bands we have the following linearised contribution to the free energy 
\begin{equation}
 F_{lin}^{con} = \frac{1}{2}\sum_{\alpha=1}^n \left(\gamma_{\alpha i j}^{-1} \partial_j\epsilon_\alpha\gamma_{\alpha i k}^{-1} \partial_k\epsilon_\alpha + \mu_\alpha^2 \epsilon_\alpha^2 \right)
 \label{linea}
\end{equation}

\noindent where $\mu^2_\alpha = \frac{1}{2}\Gamma_\alpha {\psi^0_\alpha}^2$.
 {From this linearized theory one can extract long-range core-core contribution to intervortex forces by generalising the procedure from \cite{speight97,Carlstrom.Babaev.ea:11} to the multicomponent anisotropic case. Namely we first find the equations of motion from the contribution to the free energy from equation \ref{linea}. We then spatially rescale each of these equations such that $r \rightarrow \rescaler_\alpha$ for the equation dependent on $\epsilon_\alpha$. Note that $\epsilon_\alpha$ is only dependent on this quantity due to our approximation presented in \ref{Eq:ansatzVort}. We then wish to replicate the asymptotics of the vortex scalar fields in the linear system above by including a point source $\rho \psi$ such that our system of decoupled linearised equations become the familiar wave equations,

\begin{equation}
\left(\square_\alpha + \mu_\alpha^2\right)\left|\psi_\alpha\right| = \rho_\alpha
\end{equation} 
where $\square_\alpha$ is the rescaled d'Alembert operator such that $x\rightarrow \gamma_{\alpha x} x$ and $y \rightarrow \gamma_{\alpha y}y $ or $r \rightarrow \rescaler_\alpha = \sqrt{\gamma_{\alpha x}^2 x^2 + \gamma_{\alpha y}^2 y^2}$. We have thus acquired multiple decoupled wave equations that are of the form of that considered in \cite{speight97}.}
 We follow the procedure presented in detail there for solving this equation and finding the corresponding interaction energy.  We will not reproduce the details here, see Refs \cite{speight97,Carlstrom.Babaev.ea:11} for further detail. This will lead to a point source of the form $\rho_\alpha = q_\alpha\delta(x)$ and yields the long-range interaction energy in the form of modified bessel functions $K_0$ that exponentially decay at the coherence length scales (which are the inverse masses $\mu_\alpha$ of the fields of the above linearised theory). The resulting core-core interaction energy is,
 \begin{equation}
 E_{int}^{core-core} = - \frac{q_1^2}{2\pi}
 K_0\left(\mu_1 \rescaler_1\right) - \frac{q_2^2}{2\pi}
 K_0\left(\mu_2 \rescaler_2\right)
 \label{Eq:Elin1}
 \end{equation}
  where $K_0$ is the Macdonald function. 
{ The core-core interaction is attractive and contains multiple coherence lengths, as is already understood for multi-band models. Adding the interaction which originates from magnetic and current-current forces, calculated in \cite{silaev2018non}, gives a complete picture of the long-range forces between the composite vortices:}
\begin{equation}
E_{int} = \frac{m_1^2}{2\pi}K_0\left(\lambda_L^{-1} r\right) - \frac{q_1^2}{2\pi}K_0\left(\mu_1 \rescaler_1\right) - \frac{q_2^2}{2\pi}K_0\left(\mu_2 \rescaler_2\right)
\label{Eq:Elinc}
\end{equation}
{ The first term in Eq.(\ref{Eq:Elinc}) represents the linear interaction of the magnetic origin that has the range of the London penetration length, given from \ref{Eq:kSolution} $\lambda_L^{-1} = -i k_1^{-1}$. In the limit of vanishing Josephson coupling there is only one magnetic length scale (see detail in \cite{silaev2018non}): }
\begin{equation}
k_1 = \frac{i\sqrt{\hat\gamma_{1y}^{-2}\hat\gamma_{2y}^{-2}\hat\gamma_{Lx}^{-2} \cos^2{\theta} + \hat\gamma_{1x}^{-2}\hat\gamma_{2x}^{-2}\hat\gamma_{Ly}^{-2}\sin^2{\theta}}}{\sqrt{\left(\hat\gamma^{-2}_{1y}\cos^2{\theta} + \hat\gamma_{1x}^{-2}\sin^2{\theta}\right)\left(\hat\gamma_{2y}^{-2}\cos^2{\theta} + \hat\gamma_{2x}^{-2}\sin^2{\theta}\right)}}
\end{equation} 
 {  We stress that }in general (when the Hessian is not diagonalised) these interaction terms can not be directly associated with a particular band, but are formed from combinations of parameters from all the bands. The condensate interaction terms are attractive, however with coherence lengths being rescaled by the anisotropy and hence are directionally dependent. The magnetic field however is repulsive, with it's directional dependence given by the anisotropies in the two condensates. If we reintroduced the Josephson term however we would observe an additional magnetic/current-current interaction contribution, associated with a new magnetic mode \cite{silaev2018non}, with a more complex directional dependence, and also a non-trivial form for $\mu_\alpha^2$.

We now consider the possible {  fundamental lengths scale} hierarchies for a two band superconductor with no Josephson inter-band coupling and general anisotropy. For this purpose we analyse  the interaction energy between two vortices separated by a given distance along the $x$ and $y$ directions.  
 The above parameters $\mu_\alpha$ give the exponential decay of a small perturbation in the modulus of the complex superconducting field components and thus coherence lengths: { (cf the analysis in the isotropic case  \cite{Babaev.Carlstrom.ea:10,Carlstrom.Babaev.ea:11},}
 \begin{equation}
 \xi_{\alpha i}= \mu^{-1}_\alpha \gamma_{\alpha i} = \Gamma_\alpha^{-1} 
 u_\alpha^{-1} \hat\gamma_{\alpha i}
 \end{equation}
As we are purely interested in the difference between the coherence lengths in the two directions and not the { absolute } scales, we can choose an arbitrary spatial rescaling without loss of generality. It is easiest to work with $\hat\gamma_{Lx} = \hat\gamma_{Ly} = 1$, such that the magnetic field penetration length becomes $1$ in both the $x$ and $y$ directions. This leads to the following relation $\hat\gamma_{2x}^{-2} = 1 - \hat\gamma_{1x}^{-2}$ and thus for the two band case, leaves us with effectively 4 parameter choices $\mu_1$, $\mu_2$, $\gamma_{1x}$ and $\gamma_{1y}$ such that our correlation lengths are $\xi_{1 i} = \mu_{1} \gamma_{1i}$ and $\xi_{2i} = \mu_2 \gamma_{2i}$. Finally this leads to the following interaction energies,
\begin{widetext}
\begin{eqnarray}
E_{int}^x &=& \frac{m_1^2}{2\pi}K_0\left(x\right) - \frac{q_1^2}{2\pi}K_0\left(\xi_{1x}^{-1} x\right) - \frac{q_2^2}{2\pi}K_0\left(\xi_{2x}^{-1} x\right) \\
E_{int}^y &=& \frac{m_1^2}{2\pi}K_0\left(y\right) - \frac{q_1^2}{2\pi}K_0\left(\xi_{1y}^{-1} y\right) - \frac{q_2^2}{2\pi}K_0\left(\xi_{2y}^{-1} y\right)\label{Eq:seperated}
\end{eqnarray}
\end{widetext}

At large radial distance, the interaction energy $E_{int}$ is dominated by the mode with the longest coherence/penetration length. In \ref{Eq:seperated} the magnetic field penetration length has been rescaled by a spatial rescaling such that $\lambda_L^{-1} = I$ and with loss of generality set $u_\alpha = 1$ as any change in $u_\alpha$ can be absorbed into $\Gamma_\alpha$ and $\gamma_\alpha$. For the condensate correlation lengths $\xi_{\alpha i}$, are directionally dependent. {  The new feature that appears in the 
system in question is that they can be longer or shorter than the magnetic field penetration length in various directions, }leading to different hierarchies in one direction as opposed to another. We outline below that even with no interband coupling, there are new regimes in the parameter space where the hierarchies are a mixture of other more familiar types, that do not exist in the isotropic case,
\begin{itemize}
\item \emph{Type-1/Type-1.5} - in the $x$-direction both coherence lengths are larger than the penetration length of the magnetic field $\mu_1^{-1} > \gamma_{1x}$, $\mu_2^{-1} > \left(1-\gamma_{1x}^{-2}\right)^{-\frac{1}{2}}$ and in the $y$-direction there is a mixture of hierarchies $\mu_1^{-1} > \gamma_{1y}$, $\mu_2^{-1} < \left(1-\gamma_{1y}^{-2}\right)^{-\frac{1}{2}}$.
\item \emph{Type-2/Type-1.5} - in the $x$-direction both coherence lengths are smaller than the penetration length of the magnetic field $\mu_1^{-1} < \gamma_{1x}$, $\mu_2^{-1} < \left(1-\gamma_{1x}^{-2}\right)^{-\frac{1}{2}}$ and in the $y$-direction there is a mixture of hierarchies $\mu_1^{-1} > \gamma_{1y}$, $\mu_2^{-1} < \left(1-\gamma_{1y}^{-2}\right)^{-\frac{1}{2}}$.
\item \emph{Type-1.5/Type-1.5} - in the $x$-direction there is a mixture of hierarchies $\mu_1^{-1} > \gamma_{1x}$, $\mu_2^{-1} < 1-\gamma_{1x}$ and in the $y$-direction there is also a mixture of hierarchies but in the opposite order $\mu_1^{-1} < \gamma_{1y}$, $\mu_2^{-1} > 1-\gamma_{1y}$.
\end{itemize} 
The type 1.5/type 1.5 may look similar to the isotropic type 1.5 multicomponent system, however the dominant interaction in each band switches between the $x$ and $y$ direction {  so the structure of vortex clusters should not be expected to be the same}. One may note that Type 1/Type 2 is not featured above, this is due to not having enough parameters, as in the same formalism it requires that in the $x$-direction $\mu_1^{-1} > \gamma_{1x}$, $\mu_2^{-1} > \left(1-\gamma_{1x}^{-2}\right)^{-\frac{1}{2}}$ and in the $y$-direction $\mu_1^{-1} < \gamma_{1y}$, $\mu_2^{-1} < \left(1-\gamma_{1y}^{-2}\right)^{-\frac{1}{2}}$. But when combining these inequalities it leads to $\gamma_{1x}^{-1} < \gamma_{1y}^{-1}$ and $-\gamma_{1x}^{-1} < -\gamma_{1y}^{-1}$ which is a contradiction. Note that this statement applies only to the simplest Josephson-decoupled model, but clearly that hierarchy of the length scales is possible in more general models.

We note that the addition of the Josephson coupling above would {   lead to 
hybridization of the Leggett's and London's modes \cite{silaev2018non} and thus to an additional penetration length for the magnetic field, which has to be taken into account when considering inter-vortex interactions. This leads to non-monotonic field behaviour with field inversion at some distance from the vortex center, just as in the London model }\cite{silaev2018non}. For this it is necessary and sufficient to satisfy two conditions: $h_{2}<0$ and $k_1>k_2$, so that the magnetic mode with negative amplitude can become dominating at some distance from the vortex. {  Importantly,  }the Josephson term causes $Q_{ij} \neq 0$ in equation \ref{conlin}. The resulting mixed-gradient terms will lead to non-trivial inter-vortex interactions.

Finally we reiterate the key point of this section, that unlike in isotropic multicomponent systems, it is no longer sufficient to consider the form of $F_p$ alone to determine the long-range interactions, the crystal anisotropies must be also be taken into account to determine what type a system is and different hierarchies of the length scales in different directions become possible.

\section{Multi-Vortex Solutions}
In the previous section we demonstrated the interactions between vortices in the Ginzburg-Landau model
of anisotropic superconductor are different from that obtained in the isotropic case { and also, in general, different from that 
in the London model \cite{silaev2018non}.} These interactions are characterised by different invervortex forces in different directions. We must also consider the effect of magnetic field inversion caused by the Leggett mode. We already know that the magnetic field inversion can produce weakly bound vortex states even in the strongly type-2 region of parameter space \cite{silaev2018non}. Here we are interested in a different kind of multi-soliton or multi-quanta solution, that can be formed at different length scales due to core-core interactions. To find such bound states, we consider numerical solutions to the Ginzburg-Landau energy functional with winding $N>1$. All the parameters we have selected are not well approximated by the London limit, considered in the previous paper \cite{silaev2018non}. The numerical scheme used is explained in section III.
  
  \begin{figure}[tb!]
 \includegraphics[width=1.0\linewidth]{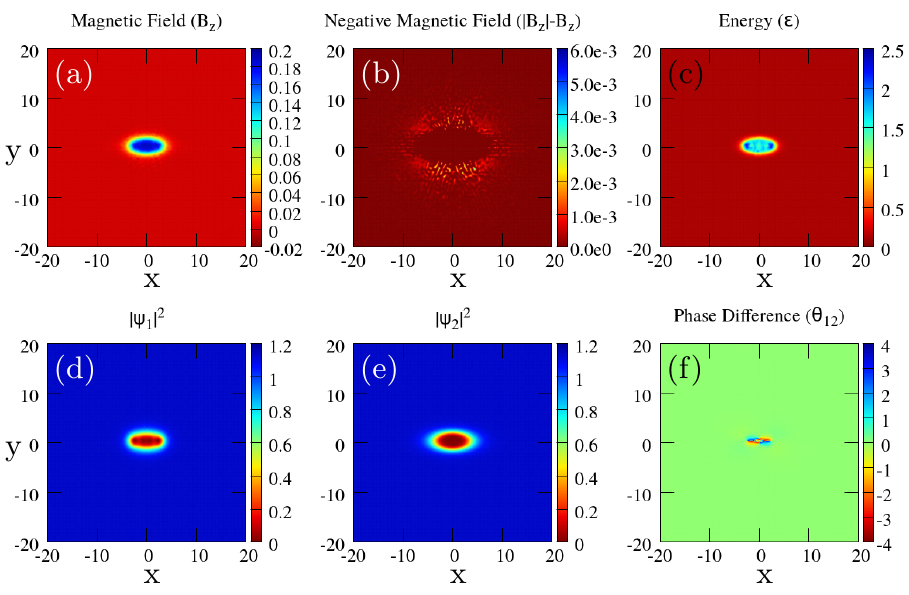}    
 \caption{\label{Fig:charge4close} (Colour online)
 $N=4$ four quanta numerical solution for anisotropy in one band exhibiting a chain  solution 
 $\Gamma_1=\Gamma_2=2$, $\eta_{12}=0.5$, $\gamma^{-1}_{2x}=2, \gamma^{-1}_{2y}=\sqrt{0.7}$ and 
 $\gamma_{1x}=\gamma_{1y} = 1$. 
 (a) $B_z$ magnetic field, (b) $\left|B_z\right| - B_z$ negative magnetic field, 
 (c) $\mathcal{E}$ energy density, 
 (d) $\left|\psi_1\right|^2$, (e) $\left|\psi_2\right|^2$,
  (f) $\theta_{12}$ phase difference.}
 \end{figure}

  \begin{figure}[tb!]
 \includegraphics[width=1.0\linewidth]{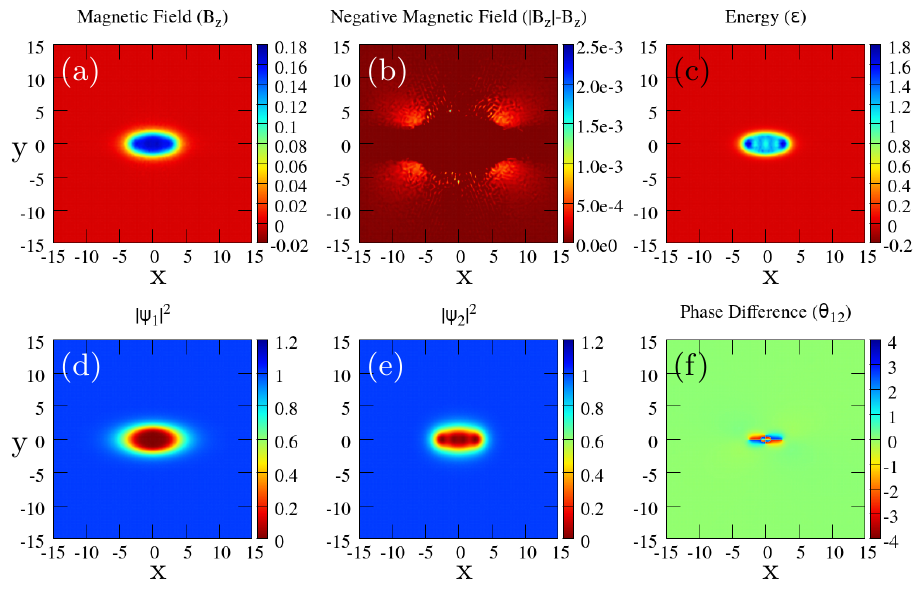}    
 \caption{\label{Fig:charge4sep} (Colour online) 
 $N=4$ four quanta numerical solution for anisotropy in one band with split parameters across the two bands allowing a clear chain to form. 
 $\Gamma_1=4, \Gamma_2=2$, $\eta_{12}=0.5$, $\gamma^{-1}_{2x}=2, \gamma^{-1}_{2y}=\sqrt{0.7}$ and 
 $\gamma_{1x}=\gamma_{1y} = 1$.
  (a) $B_z$ magnetic field, (b) $\left|B_z\right| - B_z$ negative magnetic field, 
  (c) $\mathcal{E}$ energy density, 
  (d) $\left|\psi_1\right|^2$, (e) $\left|\psi_2\right|^2$,
   (f) $\theta_{12}$ phase difference.}
 \end{figure}
   
  \begin{figure}[tb!]
 \includegraphics[width=1.0\linewidth]{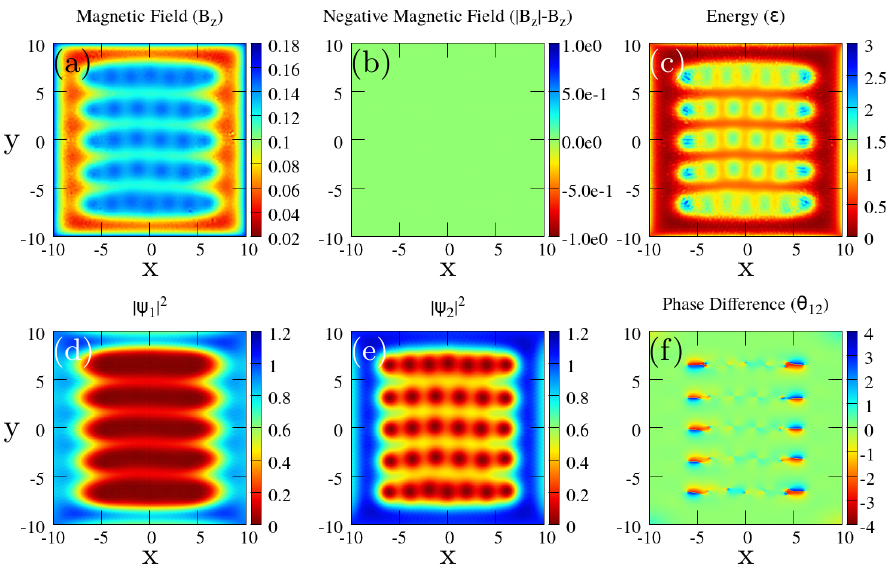}    
 \caption{\label{Fig:charge4nok} (Colour online) 
 $N=4$ four quanta numerical solution for anisotropy in one band with no Josephson coupling, allowing the chain to breathe out $\Gamma_1=\Gamma_2=2$, $\eta_{12}=0$, $\gamma^{-1}_{1x}=2, \gamma^{-1}_{1y}=\sqrt{0.7}$ and 
 $\gamma_{2x}=\gamma_{2y} = 1$.
  (a) $B_z$ magnetic field,
   (b) $\left|B_z\right| - B_z$ negative magnetic field,
    (c) $\mathcal{E}$ energy density,
     (d) $\left|\psi_1\right|^2$, (e) $\left|\psi_2\right|^2$, 
     (f) $\theta_{12}$ phase difference.}
 \end{figure}
 

We have considered a number of ($N=4$) four quanta solutions with a selection of parameters plotted in figures \ref{Fig:charge4close},\ref{Fig:charge4sep},\ref{Fig:charge4nok} that demonstrate chain-like solutions. These are the multi-quanta solutions obtained in different regimes discussed in the previous section. Namely, selecting the parameters such that a different correlation length dominates in the $x$ and $y$ direction.

In figure \ref{Fig:charge4close} we show the distributions of fields
generated by the the cluster of four vortices in the superconductor whoch has type 1.5 behaviour  in the $x$ direction and the type 2 one in the $y$ direction. The bound states take the form of chains which increase in length as the winding number increases. The chain looks similar to a 1.5 type solution with visibly separated fractional vortices in the first component but with the zeros very close. { The binding of vortices is due to the attractive core-core interaction
which dominates in the x-direction. Note that the magnetic field inversion occurs at much larger distances, meaning the field inversion has a much weaker contribution to the interaction.} We also observe that the separation of the solitons on the end of the chain is higher and with smaller solitons. Additionally the energy shown in figure \ref{Fig:charge4close}(c) has peaks at the ends of the chain.
 
In figure \ref{Fig:charge4sep} we show a similar system that is type 1.5/ type-2 but with one of the modes  { being closer to type-2 regime, so} that the vortices are more separated. Note that for this choice of parameters the additional separation and energy on the ends of the chain is less noticeable. { Figure \ref{Fig:charge4sep} shows the parameter set which is closer to type-1 behaviour. }

The binding of vortices into the chain is due to core-core interaction and the existence of long coherence length in the horizontal direction. At the same time existence of a short coherence length prevents megavortex formation: clearly there are spatially separated core singularities for both parameters. If a vortex is placed above or below a chain but very close such that non-linear effects are at play, it will be pulled into the chain, forcing other vortices out of it's way.

If other chains or vortices are placed with a large $y$ separation then they repel, due to the magnetic field penetration length being the largest length scale in this direction and hence dominating at long range. Ultimately they will form a bound state based on the negative magnetic field (the field effect considered in \cite{silaev2018non}) however as shown in the figures this is very weak, so any bound state will have negligible binding energy. 

We interpret the above solutions as having type-2 behaviour in the $y$-direction and  type-1.5 in the $x$-direction respectively. 
The new features here are (i) the separation of bound vortex cores forms in one direction rather than forming compact clusters (the consequence of the different length scale hierarchies in different directions), and (ii) type-2 like behaviour in the direction perpendicular to the chain.

\section{Magnetization}
The above demonstrated different hierarchies of the length scales in different directions,
as well as the structure of isolated solutions.
 We are now interested in { the question of} how these chain/stripe solutions enter into a magnetised sample in experiment. To model the magnetisation of a finite domain or sample we must introduce the external field $H$ { (in the previous sections the vortices were created by an initial guess and stayed in the sample because of negligible interaction with the boundary)}.  Hence in this section we minimize the Gibbs free energy on a finite domain $G = F - 2\int_{\mathbb{R}^2} H\cdot B \; d^2 x$. We impose the condition $\nabla\times \boldsymbol{A} = H$ on the boundary of the superconductor. If we then slowly increase the external field value in steps of $10^{-2}$ we can simulate the turning up of an external field and the subsequent magnetisation of the theory over our finite domain.

\begin{figure}[tb!]
 \includegraphics[width=1.0\linewidth]{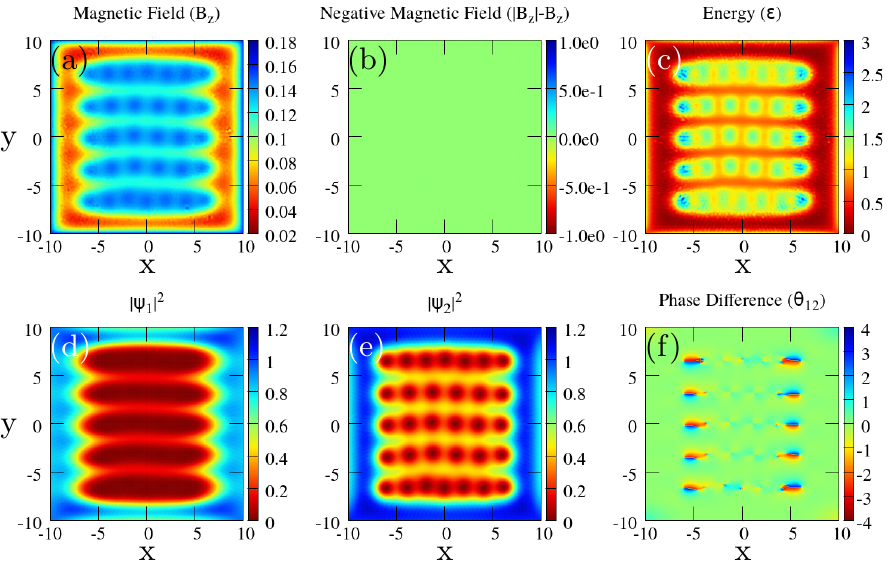}    
 \caption{\label{Fig:mag1} (Colour online)
 Magnetisation  numerical solution for anisotropy in one band $\gamma_{1x} = 0.7$, $\gamma_{1y} = 0.4\gamma$, $\gamma_{2x} = \gamma_{2y}$, $\eta_{12} = 0.5$ and $\Gamma_{1}=\Gamma_{2}=2$. 
 The contour plots are (a) $B_z$ magnetic field, (b) $\left|B_z\right| - B_z$ negative magnetic field, 
 (c) $\mathcal{E}$ energy density, 
 (d) $\left|\psi_1\right|^2$, (e) $\left|\psi_2\right|^2$, 
  (f) $\theta_{12}$ phase difference. 
  The picture shows a clear tendency for vortex stripes formation with the stripes being vortex bound states due to core-core interaction.}
 \end{figure}

The results of such simulations for the parameters $\gamma_{1x}^{-1} = \sqrt{4.5}$, $\gamma_{1y}^{-1} = \sqrt{0.7}$, $\gamma_{1x}^{-1} = \gamma_{1y}^{-1} = 1$, $\Gamma_{1} = \Gamma_{2} = 2$ and $\eta_{12} = 0.5$ are shown in figure \ref{Fig:mag1}. If we consider the results here we can see that the vortices are interacting in very different ways in the $x$ and $y$ directions in the domain. This causes vortex chains to form. The entire domain from the simulation is plotted and only one value of external magnetic field is shown. The magnetization process   { has the form of a first order phase transition associated with the sudden entry of a large number of vortices due to attractive core-core interaction}.

Analysing how the chains interact is challenging and we must use an alternative method. 

\section{Interaction  of vortex chains}
In this section we are interested in demonstrating that chain-like solutions do form in the bulk of superconductors subject to an external magnetic field. Also we discuss how such objects are formed and interact with each other. We have already demonstrated that a regime exists in parameter space where vortex solutions in an anisotropic multiband model tend to form bound states in some direction while repelling in other direction. However this does not necessarily mean that clear well-separated chains can form in a magnetised sample. To study this we must study the chain solutions themselves, which is easiest to achieve by studying vortex solutions in a periodic space.

The easiest way to achieve this is to consider a domain that is periodic in one direction (namely the direction that the stripes/chains prefer to form). We have therefore introduced periodic boundary conditions in this direction, such that $\psi_\alpha (L, y) = \psi_\alpha( 0 , y)$ and $A(L,y) = A(0,y)$. The other two boundary conditions are then just the standard zero normal-current condition. These boundary conditions  allow the correct winding for map to occur and thus for vortex solutions to traverse the space. All the winding will be located on the boundaries with zero normal-current condition due to the periodic conditions.

\begin{figure*}[tb!]
 \includegraphics[width=1.0\linewidth]{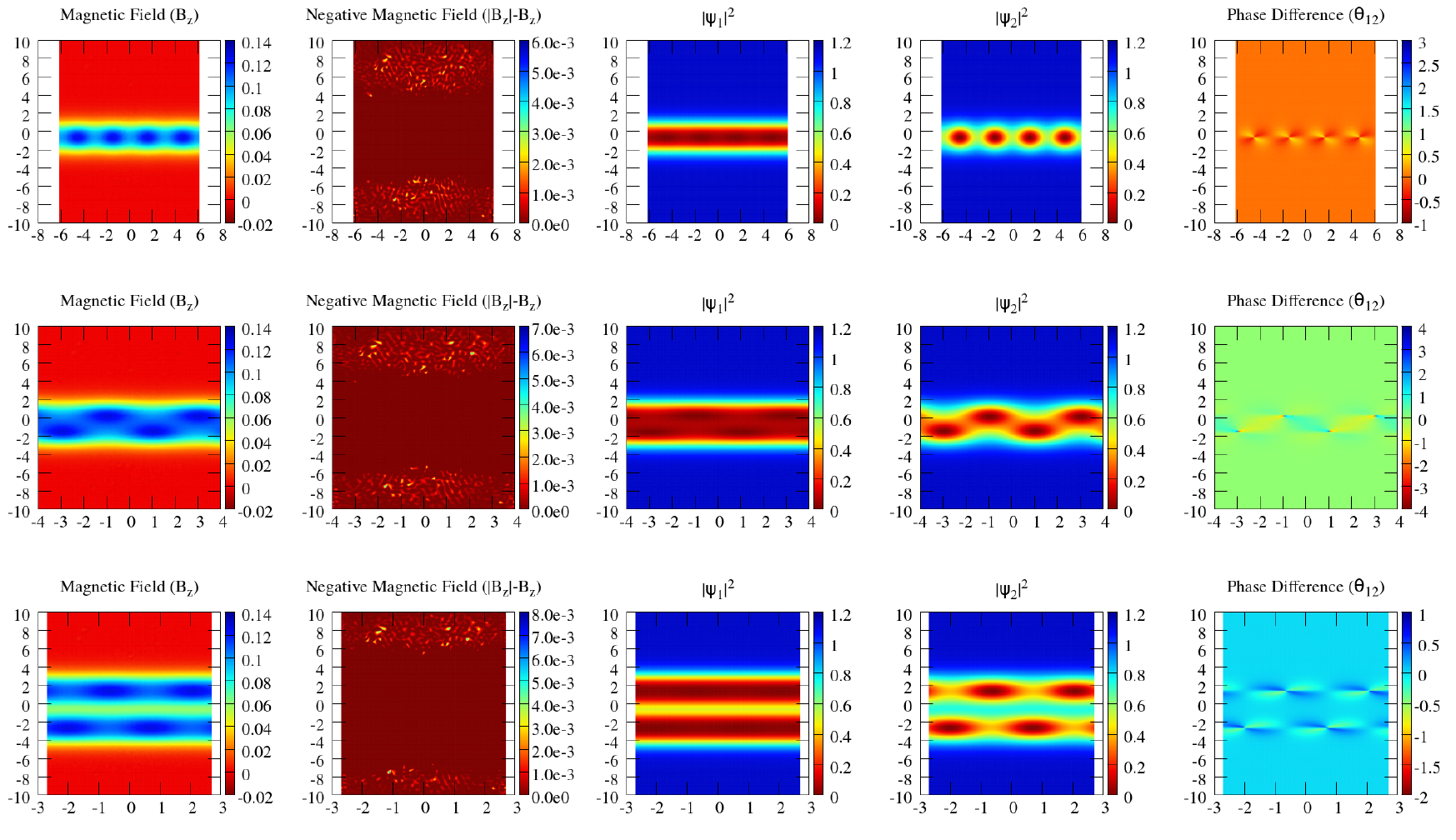}    
 \caption{\label{Fig:periodic1} $N=4$ four quanta global minimal energy solutions, {with periodicity enforced in the horizontal direction,} is successively squeezed (periodic length is shortened). This squeezing results in an effective increase in magnetic field density. Note that the Josephson coupling is switched on here, which has a marked effect on the length scales. We observe the chain buckling and then splitting, with the chains having an optimal separation which remains stable for long run times. The parameters are $\Gamma_1 = \Gamma_2 = 2$ $\gamma_{1x}^{-2} = 4.5$, $\gamma_{1y}^{-2} = 0.7$, $\gamma_{2x} = \gamma_{2y} = 1$ and $k_0 = 1$. } 
 \end{figure*}

\begin{figure*}[tb!]
 \includegraphics[width=1.0\linewidth]{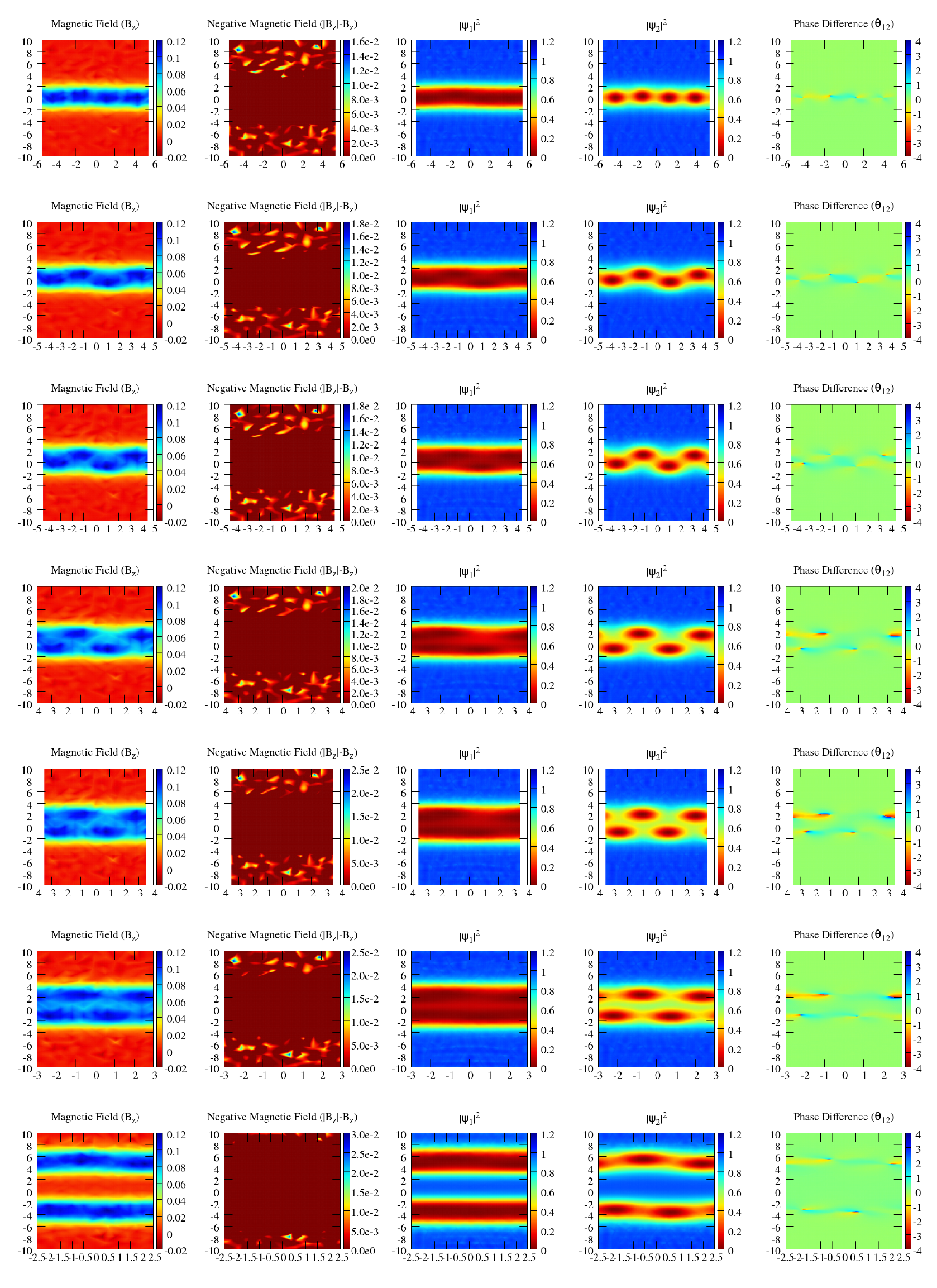}    
 \caption{The plots show the minimal energy solutions for a space that is periodic in { the horizontal} direction where the periodic length $L$ is continuously decreased, representing an effective increase in magnetic field density. Note that the Josephson term has been switched off here. The chain starts to buckle and eventually splits into two chains. These two chains ultimately repel each other becoming well seperated. This demonstrate direction-depdent typology of the superconducting state: it is type-1.5-like vortex bound states formation in x-direction (dominated by core-core interaction) and type-2-like vortex chain repulsion in y-direction (dominated by current-current and magnetic interactions).The parameters are $\Gamma_1 = \Gamma_2 = 2$ $\gamma_{1x}^{-2} = 4$, $\gamma_{1y}^{-2} = 0.7$, $\gamma_{2x} = \gamma_{2y} = 1$ and $k_0 = 0$.\label{Fig:periodic2}}
 \end{figure*}

We then introduce a 4-quanta solution into the domain and track what happens as $L$ is varied. Specifically we are interested in what happens as $L$ is reduced, effectively squashing our 4-quanta solution and increasing the magnetic field density. This is achieved by using a conjugate gradient flow method as detailed in section III to minimise the configuration, then reducing $L$ and minimising again. This process is repeated until the desired magnetic field density is achieved. The field configurations for applying this method are plotted for various parameters in figures \ref{Fig:periodic1} and \ref{Fig:periodic2}.

There are a few different categories that the various parameters which give chain like solutions can fall into.
 Firstly as the periodic length $L$ is reduced
the stripe homogenises in the $x$-direction and then continues to expand into the $y$-direction, eventually filling the space with the homogeneous value $\left|\psi\right|_{\alpha}^2 = 0$.

All the other categories that solutions can fall into exhibit some form of chain splitting as the magnetic field density is increased. In figure \ref{Fig:periodic1} we can see that the familiar chain solution initially forms and as it is squashed it buckles forming a zig-zag and then splits into multiple chains. These multiple chains still have some separation in the zeroes of the second condensate which interlace with the zeroes in the other chain. Note that these are the same parameters as for the magnetisation shown in figure \ref{Fig:mag1}.

In figure \ref{Fig:periodic2} we see a similar effect in that the chain starts to buckle and splits. However we now see as the magnetic field density increases that the individual chains repel each other in a stronger fashion. This leads ultimately to two well separated stripes. In that case we have the type-2 magnetic-dominated repulsive interaction in vertical direction and type-1.5 core-core-interaction-dominated intervortex forces in x-direction. Note however that similarly to the extreme type-2 case \cite{silaev2018non} in such a regime there can appear a small attractive force between stripes due to field inversion at a wide separation and hence not an infinite optimal separation.

This behaviour is the best demonstration of the vortices acting very differently in different directions in the domain. Attracting vortices in one direction onto the end of the chains and repelling vortices in the other direction.

\section{Conclusion}

In conclusion, we have discussed generalisation of the length-scale-hierarchy-based typology 
of superconductors  \cite{landau1950k,ginzburg1952} to the case where both multiple components and anisotropies are present.  In an isotropic multicomponent case, the earlier discussed regimes were
type-1 where all coherence lengths are larger than the magnetic field penetration length $\xi_i>\lambda$, type-2 where $\lambda<\xi_i$ and a mixed case where some of the coherence lengths are larger and some are smaller than the magnetic field penetration lengths (that was termed earlier ``type-1.5"). Besides that one should distinguish the special { zero-measure Sarma-Kramer-Bogomolny critical point, where all these length scales are equal and the Bogomolny bound is saturated. We have demonstrated that in the multi-component case the length-scale-based typology is quite different from the anisotropic single-component counterpart \cite{ginzburg1952,kats1969some}.
 The considered anisotropic $n$-band multicomponent system is characterized by $n$-coherence lengths that differ in different directions, $\xi_1(\hat{{\bf r}}),\xi_2(\hat{{\bf r}}),...,\xi_n(\hat{{\bf r}})$ as well as multiple directionally dependent magnetic field penetration lengths $\lambda_1(\hat{{\bf r}}),\lambda_2(\hat{{\bf r}}),...\lambda_{n+1}(\hat{{\bf r}})$ (of which there are $n+1$ when the Leggett and magnetic modes are hybridized \cite{silaev2018non}).
  The new feature that arises is that the hierarchies of these length scales are different for different directions in a crystal.}
 From the obtained asymptotic intervortex forces it is clear that vortex structure formation in these systems should in general be quite complicated and warrants further study. A new regime that we discussed in particular is where, for a magnetic field 
applied in the $z$ direction, the length scale hierarchy is type-1.5 in the $x$-direction and type-2 in $y$ direction. In that regime the system forms vortex bound states in the form of stripes or chains, where the intervortex attraction is mediated by core-core overlap in $x$-direction while the stripes have type-2 interaction in the $y$ direction, mediated by the magnetic and current-current forces. In the language of interface energies, the system tries to maximize interfaces in the $y$ direction while in $x$-direction there are different interfaces, some of which the system tries to maximize and some that the system tries to minimize (i.e. in some of the regime the components clearly forms a ``mega-vortex-like" core extending in $x$-direction). By the same token the non-re-scalability of length scale hierarchies suggests mixed type-1/type-2 regimes where the system goes into a stripe pattern that maximizes superconductor-to-normal interfaces in one direction, while being translationally invariant in a perpendicular direction. In Ref. \cite{silaev2018non} we gave some simple estimates that disparity of magnetic field penetration lengths should arise for realistic multiband materials.
Besides crystalline anisotropies, the effect of vortex chains formation due to length scale anisotropies can arise in muliband system with strains, that can lead to the locally type-1.5 hierarchy of the length scales.
This calls for further calculations regarding which microscopic parameters realize the above regimes in multiband materials.

\begin{acknowledgments}
\noindent TW would like to acknowledge fruitful discussions with Martin Speight. The work was supported by the Swedish Research Council Grants No. 642-2013-7837, VR2016-06122, Goran Gustafsson Foundation for Research in Natural Sciences and Medicine and EPSERC Grant No. EP/P024688/1. 
 MS was supported by the Academy of Finland (Project No. 297439). 
 Part of this work was performed at the Aspen Center for Physics, which is supported by National Science Foundation grant PHY-1607611.
The computations 
were performed on resources provided by the Swedish National 
Infrastructure for Computing (SNIC) at National Supercomputer 
Center at Link\"oping, Sweden. 
\end{acknowledgments}
\bibliography{references}
 \end{document}